\documentclass[useAMS,usenatbib]{mn2e}
\pdfoutput=1
\usepackage{rotating}
\usepackage{journals}
\usepackage{graphicx, subfigure}
\usepackage{xspace}
\usepackage{amssymb}

\def\source{GRS~1915$+$105\xspace}

\LARGE \normalsize \title{On the harmonics of the low-frequency quasi-periodic oscillation in \source}

\author[Ratti \& Belloni] {E.M.~Ratti$^1$\thanks{email : e.m.ratti@sron.nl}, T.M.~Belloni$^{2}$, S.E.~Motta$^{2}$\\
$^1$SRON, Netherlands Institute for Space Research, Sorbonnelaan 2, 3584~CA, Utrecht, The Netherlands\\ 
$^2$INAF-Osservatorio Astronomico di Brera, via E. Bianchi 46, 23807 Merate (LC), Italy \\ }

\begin{document}

\maketitle

\begin{abstract} 
\noindent
 \source is a widely  studied  black hole binary, well known because of its extremely fast and complex variability. Flaring periods of high variability alternate with ``stable'' phases
 (the {\it plateaux}) when the flux is low, the spectra are hard and the timing
 properties of the source are similar to those of a number of  black
 hole candidates in hard spectral state. In the {\it plateaux} the power density spectra are dominated
 by a low frequency quasi periodic oscillation (LFQPO) superposed onto a band limited noise continuum and accompanied by at least one harmonic.
 In this paper we focus on three  {\it plateaux}, presenting the analysis of the power density spectra and in particular of the LFQPO and its harmonics. While plotting the LFQPO and all the harmonics together on a frequency-width plane, we found the presence of a positive trend of broadening when the frequency increases. This trend can shed light in the nature of the harmonic content of the LFQPO and challenges the usual interpretation of these timing features.

\end{abstract}

\section{introduction}
\begin{figure}
 \includegraphics[width=9cm]{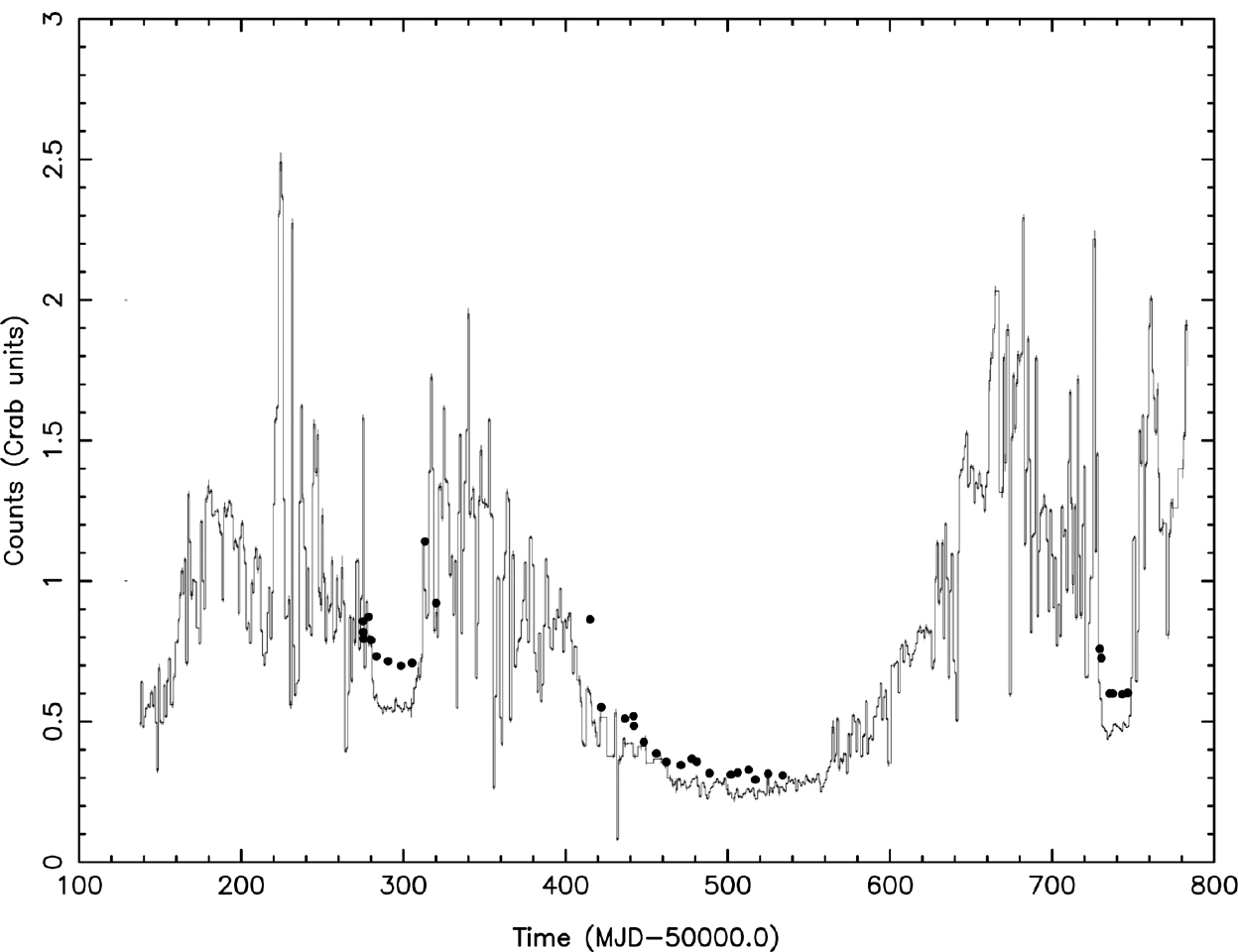}
 \caption{The {\it RXTE}/PCA observations analysed in this paper (black points) superimposed onto the {\it RXTE}/ASM daily$-$average light curve between October 1995 and December 1997 (from \citet{2000A&A...355..271B}). The flux level is different because of the different energy band of PCA and the ASM. PCA data belong to three {\it plateaux}, the middle one being significantly longer that the others.}
\label{LICU}
 \end{figure}

X-ray binaries are binary systems where a stellar compact object, a neutron star (NS) or a black hole (BH) accretes matter from a companion star, producing bright X-ray emission. Since its launch in 1995, the {\it Rossi X-Ray Timing Explorer (RXTE)} has allowed a direct comparison between the spectral and timing properties of large sample of XRBs, which allowed the  definition of the typical phenomenological properties of  sources hosting a NS or a BH (see \citealt{2006csxs.book....1P} for a review).  On the base of those properties, quite a number of sources have been identified as possibly hosting a BH (black-hole candidates, BHC): for a number of systems this prediction has been confirmed via dynamical studies. 
\newline
 Most BHCs have a low-mass companion star and are transient sources (black-hole transients, BHT) showing an evolution of their spectral/timing properties during their outbursts. This evolution has been described by many authors as a transition through a combination of 3-4 states (see \citealt{Rem06} and \citealt{Bel10} for a review) mainly defined by the shape and features in the power density spectra (PDS) and by the hardness of the energy spectra.  The  typical PDS are composed by a band-limited noise continuum that varies during the outburst  and by a number of quasi-periodic oscillations (QPOs) observed  in association with certain states both  at low (mHz to few tens of Hz) and high (tents to hundreds Hz) frequencies  \citep{2006csxs.book..39V}. 
 \newline
 Various models have been proposed in order to explain the quasi periodic variability and the noise continuum (see \citealt {2005AN....326..798V} for a review, \citealt{2011MNRAS.415.2323I}  for a more recent model) but none of them,  at the present time,  can give a comprehensive explanation of all the observed phenomena.  \newline
 In this paper we focus on the low frequency QPO (LFQPO) of \source . LFQPOs are common features in the PDS of BHTs (unlike high-frequency QPOs, with frequencies larger than $\sim$ 30 Hz, that are rather rare) during their hard spectral states. They appear as strong peaks at a centroid frequency between 0.01 and 15 Hz, with a fractional rms amplitude reaching values $>$15$\%$ in sources like XTE~J1550$-$564 and \source ( \citealt{2000ApJ...544..993S}, \citealt{2000ApJ...541..883R}). 
LFQPOs are often accompanied by  one or more further QPO peaks whose frequency is in harmonic relation with that of the LFQPO itself. Those are traditionally interpreted as resulting from the Fourier decomposition of the quasi-periodic signal responsible for the LFQPO. Nonetheless, questions about the identification of the fundamental peak and  the genuineness of the harmonic relationship of the peaks have been raised in the recent works of \citet{2010ApJ...714.1065R} and \citet{2011arXiv1104.3865R}. With the aim of  investigating the relation between the LFQPO and its harmonics, we analysed a sample of {\it RXTE} observations from \source,  which is particularly favorable for this study due to the strength of the main QPO peak and the number of the harmonics. The source is a largely studied XRB (see \citealt{2004ARA&A..42..317F} for a review) with a K-M III \citep{2001A&A...373L..37G}  star donating matter to a dynamically confirmed BH \citep{2001Natur.414..522G}.  It was discovered as a transient source on August 15, 1992 by the {\it WATCH} instrument on board on the {\it GRANAT} satellite \citep{1992IAUC.5590....2C} and since then
 it always remained bright in the X-ray sky ($\sim$ 0.5 - 2 Crab), undergoing an 18-year long outburst that at the time of writing is still ongoing.
 Starting from 1996, \source  has been monitored
 with {\it RXTE}. The light curve in Fig. \ref{LICU} is an example of the long-scale behavior
 of the source: quiet periods of low flux and low variability lasting hours to tens of days (the {\it plateaux}, \citealt{1999MNRAS.304..865F}), alternate with flaring activity phases of extreme variability, when all the X-ray properties, flux, spectral parameters
 and PDS features can vary on timescales down to the ms (\citealt{1996ApJ...473L.107G}, \citealt{1997ApJ...488L.109B}, \citealt{2000A&A...355..271B}). While peculiar and complex during the flaring phases, the behaviour of the source during the {\it plateaux} is  similar to that of many other BHTs. 
In particular, the PDS show the typical band limited noise complex (extending up to $\sim100~Hz$) and a strong LFQPO with several harmonics (\citealt{2001ApJ...558..276T}, \citealt{2000ApJ...541..883R}).
We analysed {\it RXTE} data from the three {\it plateaux} in Fig. \ref{LICU}: the data sample and the reduction method are described in Sect. \ref{data}, while the modeling of the PDS is described in Sect. \ref{mod}. Sect. \ref{harm} presents our results, which are discussed in the final Sect. \ref{disc}.

\begin{table}
\label{log}
\caption{List of the {\it RXTE}/PCA archival observations analysed in this paper. The observations are divided in three blocks, corresponding to the three subsequent {\it plateaux} in Fig. \ref{LICU}. }
\begin{center}
\begin{tabular}{ccc}
\hline
Obs & MJD & Counts \\
 &  (50000.0)& (Crab units) \\
\hline
10408$-$01$-$22$-$01 & 275.2230 & 0.818 \\
10408$-$01$-$22$-$02 & 275.3564 & 0.796 \\
10408$-$01$-$23$-$00 & 278.4915 & 0.873 \\
10408$-$01$-$24$-$00 & 280.1696 & 0.790 \\
10408$-$01$-$25$-$00 & 283.4941 & 0.732 \\
10408$-$01$-$27$-$00 & 290.5768 & 0.715 \\
10408$-$01$-$28$-$00 & 298.5349 & 0.699 \\
10408$-$01$-$29$-$00 & 305.3730 & 0.709 \\
10408$-$01$-$30$-$00 & 313.3129 & 1.141 \\
10408$-$01$-$31$-$00 & 320.1960 & 0.922 \\
\hline
20402$-$01$-$04$-$00 & 415.1299 & 0.864 \\
20402$-$01$-$05$-$00 & 421.9758 & 0.551 \\
20402$-$01$-$07$-$00 & 436.6608 & 0.511 \\
20402$-$01$-$08$-$00 & 441.9193 & 0.519 \\
20402$-$01$-$08$-$01 & 442.1189 & 0.485 \\
20402$-$01$-$09$-$00 & 448.2834 & 0.427 \\
20402$-$01$-$10$-$00 & 455.9928 & 0.387 \\
20402$-$01$-$11$-$00 & 462.0617 & 0.356 \\
20402$-$01$-$12$-$00 & 471.0688 & 0.346 \\
20402$-$01$-$13$-$00 & 477.8727 & 0.368 \\
20402$-$01$-$14$-$00 & 480.8806 & 0.357 \\
20402$-$01$-$15$-$00 & 488.7786 & 0.317 \\
20402$-$01$-$16$-$00 & 501.8847 & 0.312 \\
20402$-$01$-$18$-$00 & 512.8887 & 0.329 \\
20402$-$01$-$19$-$00 & 517.0462 & 0.294 \\
20402$-$01$-$20$-$00 & 524.9231 & 0.315 \\
20402$-$01$-$21$-$00 & 533.8351 & 0.309 \\
\hline
20402$-$01$-$49$-$00 & 729.3267 & 0.759 \\
20402$-$01$-$49$-$01 & 730.3943 & 0.726 \\
20402$-$01$-$50$-$01 & 737.4041 & 0.600 \\
20402$-$01$-$51$-$00 & 743.2941 & 0.597 \\
20402$-$01$-$52$-$00 & 746.5504 & 0.602 \\
\hline

\end{tabular}
\end{center}
\end{table}

\begin{figure}
\includegraphics[width=9cm]{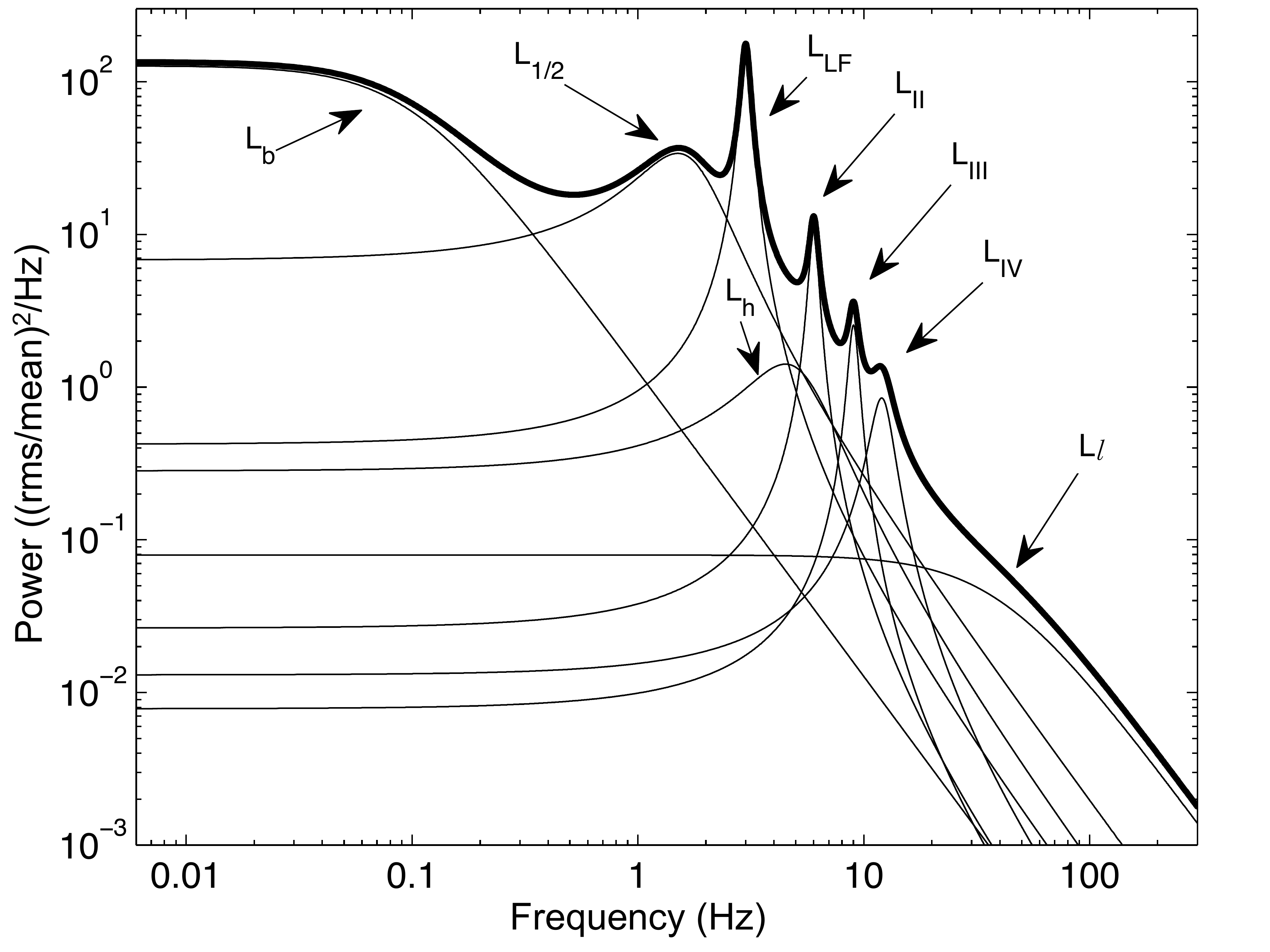}
\caption{The most general multi-Lorentzian model we applied to the PDS. The Lorentzian components are labeled according to \citet{2002ApJ...572..392B}. The L$_{IV}$ component accounts for residuals in the fit in a few observations, but is never significant.}
\label{PDSmodel}
\end{figure}

\begin{figure*}
\hbox{\hspace{0.1in} 
\includegraphics[width=7.9cm]{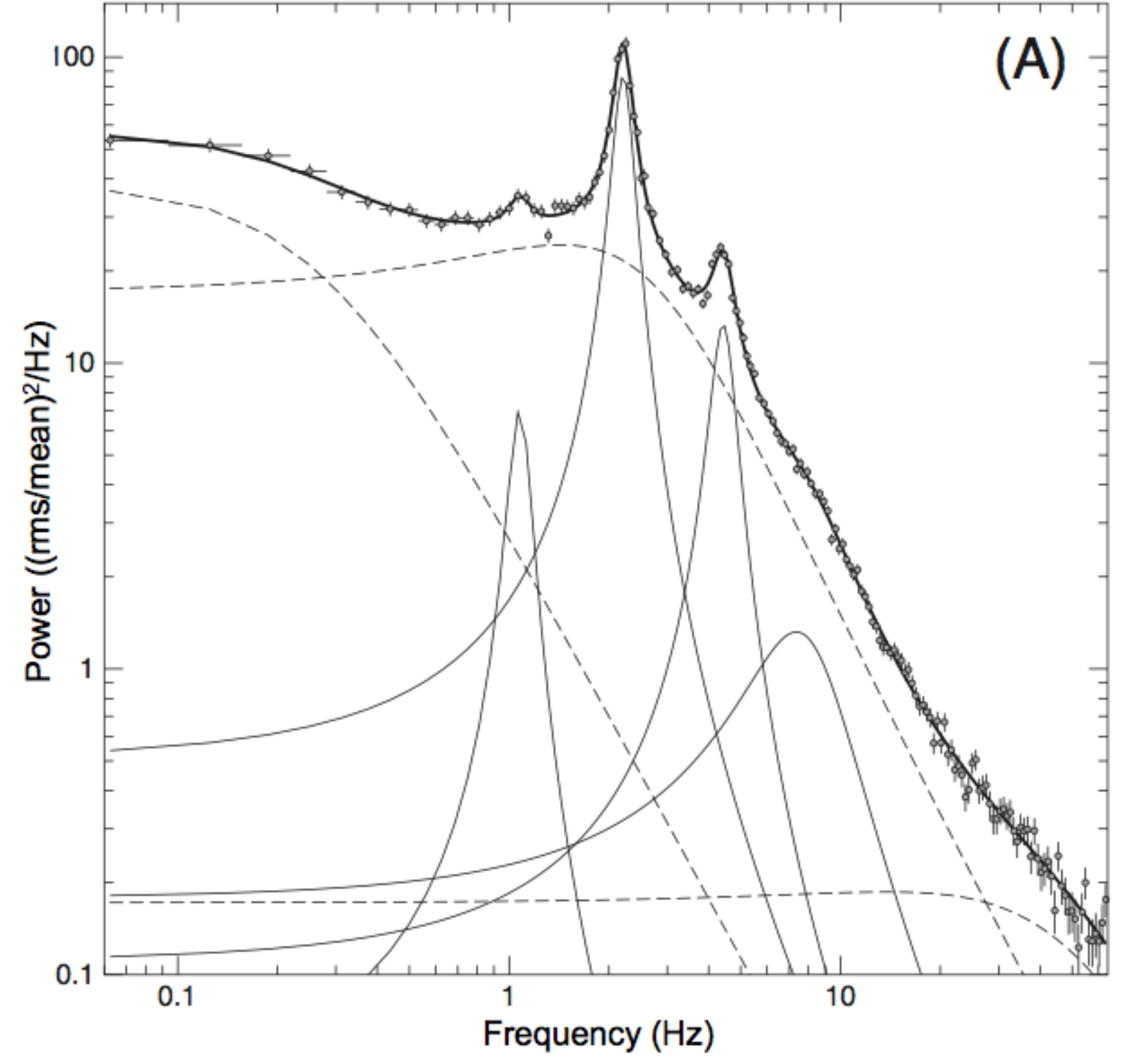}
\hspace{0.1in} 
\includegraphics[width=8cm]{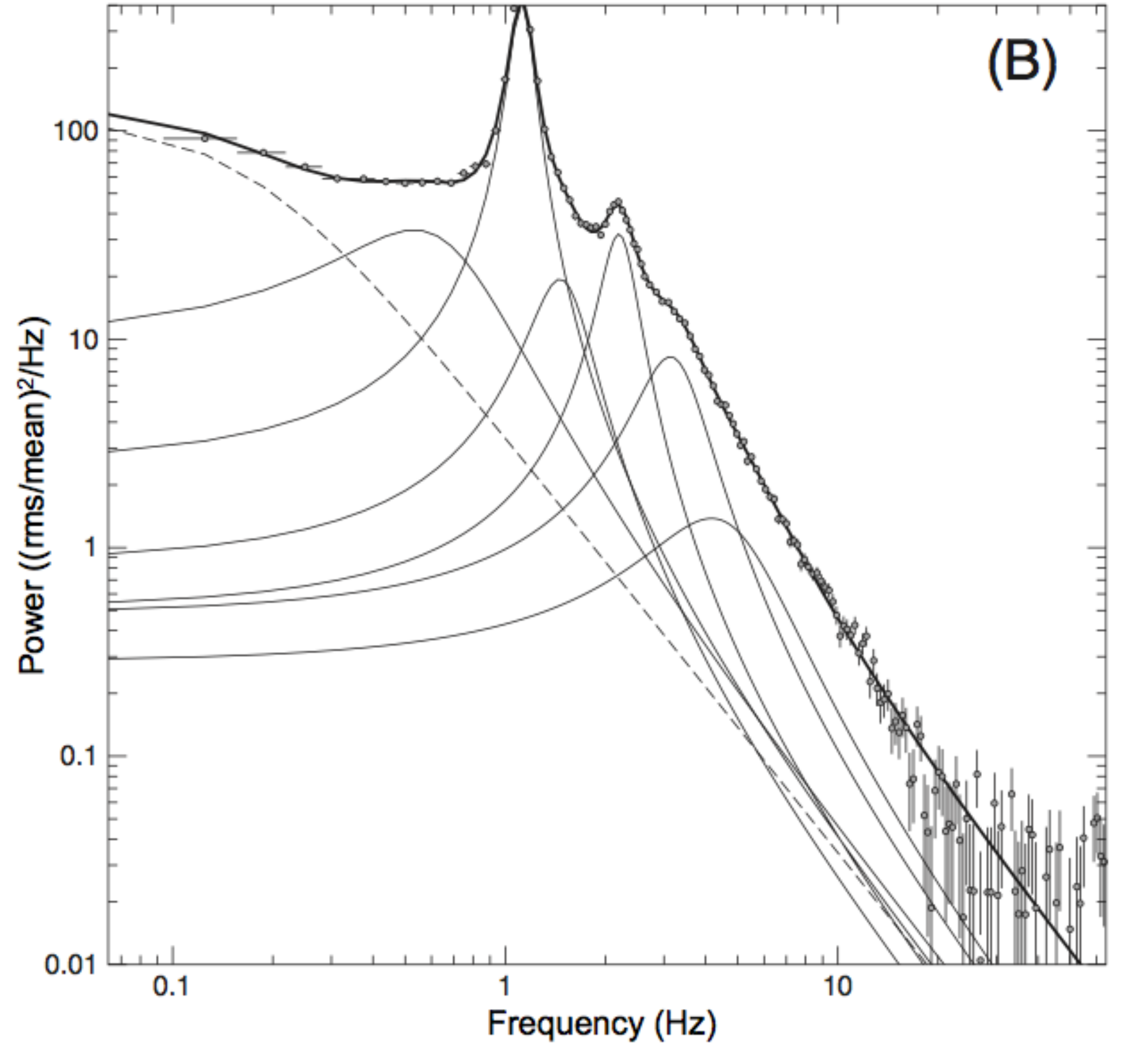}
}
\caption{PDS from observation  20402$-$01$-$19$-$00 (A) and  10408$-$01$-$27$-$00(B) with our best fitting model. 
Continuum components are plotted as dashed lines, QPOs with filled lines. 
The high frequency component $L_{\it l}$ (see Fig. \ref{PDSmodel}) that is only observed during the long {\it plateau}  \citep{2001ApJ...558..276T} can be seen in panel (A).}
\label{PDSesempi}
\end{figure*}

\section{Observations and data reduction}
\label{data}

We analysed a set of  thirty-two {\it RXTE}/PCA observations from the {\it RXTE} NASA's archive (Table \ref{log}) performed between July 1996 and October 1997 during the three subsequent {\it plateaux} shown in Fig. \ref{LICU} ( for a more general studies of the same plateaux,  \citealt{2001ApJ...558..276T}).  
\newline
Power Density Spectra (PDS) have been extracted in a $\sim2-14~\rmn{keV}$ energy band from PCA light curves (single bit data with temporal resolution 125$\mu s$) with a Nyquist frequency of $2048~Hz$,  normalised according to \citet{1983ApJ...266..160L} and converted to square fractional root mean square (rms) deviation (\citealt{1990A&A...227L..33B}).  In order to improve the statistics, we obtained a PDS for each observation as the average of  the Fourier spectra extracted from consecutive $16-s$ segments of the light curve. The averaged PDS have been re-binned logarithmically  in frequency and fitted
 with the standard {\sc Xspec} (version 11.3.2) fitting package using a diagonal response, therefore performing a simple direct $\chi^2$ minimization. We adopted a model composed by Lorentzian functions only (see Fig. \ref{PDSmodel}) following  \citet{2002ApJ...572..392B}. As argued by the authors, this approach has the advantage that the spectrum is described by  components that are directly comparable one to the other, with no assumption about the origin
 of each one. A component has been considered significant if $N/\sigma_N>3$, where $N$ is the normalisation and $\sigma_N$ its uncertainty, obtained with the chi-squared minimization. The errors on individual powers of the fitted PDS were computed following \citet{1989ARA&A..27..517V}.
We accounted for the counting statistic in PDS using an automatic subtraction of Poissonian background, based on the estimate of PCA dead time by \citet{1995ApJ...449..930Z}. A flat component extending over all the frequencies appears in a few PDS as a result of a non-perfect Poissonian subtraction. We corrected for this effect while fitting the PDS.

\section{Multi-Lorentzian modeling of the PDS}
\label{mod}

Figure \ref{PDSesempi} shows two examples from the sample of PDS we analysed  and the associated best-fitting model. Depending on the observation, five to eight Lorentzians are required for a good fit ( $\chi^2<1.4$). We interpret each Lorentzian as a different spectral component (with the exception of double peaked QPOs, see below) assigning labels as in  Fig. \ref{PDSmodel}.  We define the frequency $\nu$ and width $\Delta$ of a component as the central frequency and the FWHM of the associated Lorentzian function. 
\newline
One to three Lorentzian components account for the band limited continuum (see Fig.  \ref{PDSmodel} and \ref{PDSesempi}): a low frequency one,   $L_b$, appears in 29 over 32  observations at $\nu_b\lesssim 0.2~Hz$.  $L_h$ is often required at a centroid frequency  $\nu_h$, approximately located under the LFQPO $L_{LF}$.   
A high frequency component $L_{\it l}$  only appears in observations from the longest {\it plateau} in Fig. \ref{LICU}, at $\nu_{\it l}\gtrsim42~Hz$. For a study of this component and its connection with ``short'' and ``long'' {\it plateaux}, see \citealt {2001ApJ...558..276T}.
\newline
One Lorentzian is usually enough to fit the LFQPO peak, unless it  happens to be double-peaked. The presence of a double peak is possible because the QPO is known to drift in frequency on time scales shorted than the average observation (see e.g. \citealt{1999ApJ...513L..37M} ).
When this is the case, we fit two Lorentzians to the QPO and we take the sum-function as a single spectral component $L_{LF}$ to represent the LFQPO.  We consider the frequency of the most significant (larger $\%$rms) peak as the frequency of $L_{LF}$, $\nu_{LF}$ and define its width $\Delta_{LF}$ as the FWHM of the sum-function of the overlapping Lorentzians. The error on $\Delta_{LF}$ is analytically computed from the equation of the sum-function. The quality factor $Q=\nu/\Delta$ (typical indicator for the coherence of a signal, a QPO is traditionally defined by $Q>$2) of $L_{LF}$ is always above 3, reaching a maximum value of 10.4 (see Table \ref{tabLLF}). 
\newline
Although this is not the main topic of this paper, it is worth to mention that our data are consistent with the relations between $\nu_h$, $\nu_{\it l}$ and $\nu_{LF}$ found in \citet{2002ApJ...572..392B} for a sample of sources not including \source. 
\newline
 After fitting for the main QPO peak and the broad band-limited noise components $L_b$, $L_h$ and $L_{\it l}$, further peaks or bumps in the PDS are best fitted by up to three components in harmonic relation with $L_{LF}$, labeled $L_{II}$, $L_{III}$ and $L_{1/2}$. Their central frequencies are  $\nu_{II}\sim2\nu_{LF}$, $\nu_{III}\sim3\nu_{LF}$ and  $\nu_{1/2}\sim0.5\nu_{LF}$ respectively. In three observations a further component ($L_{IV}$ in Figure \ref{PDSmodel}) needs to be added to our model in order to take into account residuals in the fit, whose centroid frequency is consistent with $\sim4\nu_{LF}$. However, this component never results significant (slightly less than 2$\sigma$ detection) and therefore we did not consider it in our subsequent analysis. The component $L_{II}$ always appear as a rather coherent peak, with $2\lesssim Q\lesssim10$. $L_{III}$ and $L_{1/2}$ instead are often broad, with $Q$ ranging from less than 1 up to $\sim$8 (see Table \ref{tabLII}, \ref{tabLIII}, \ref{tabL1/2}).

\begin{figure}
\centering
\begin{tabular}{c}
\includegraphics[width=8.5cm] {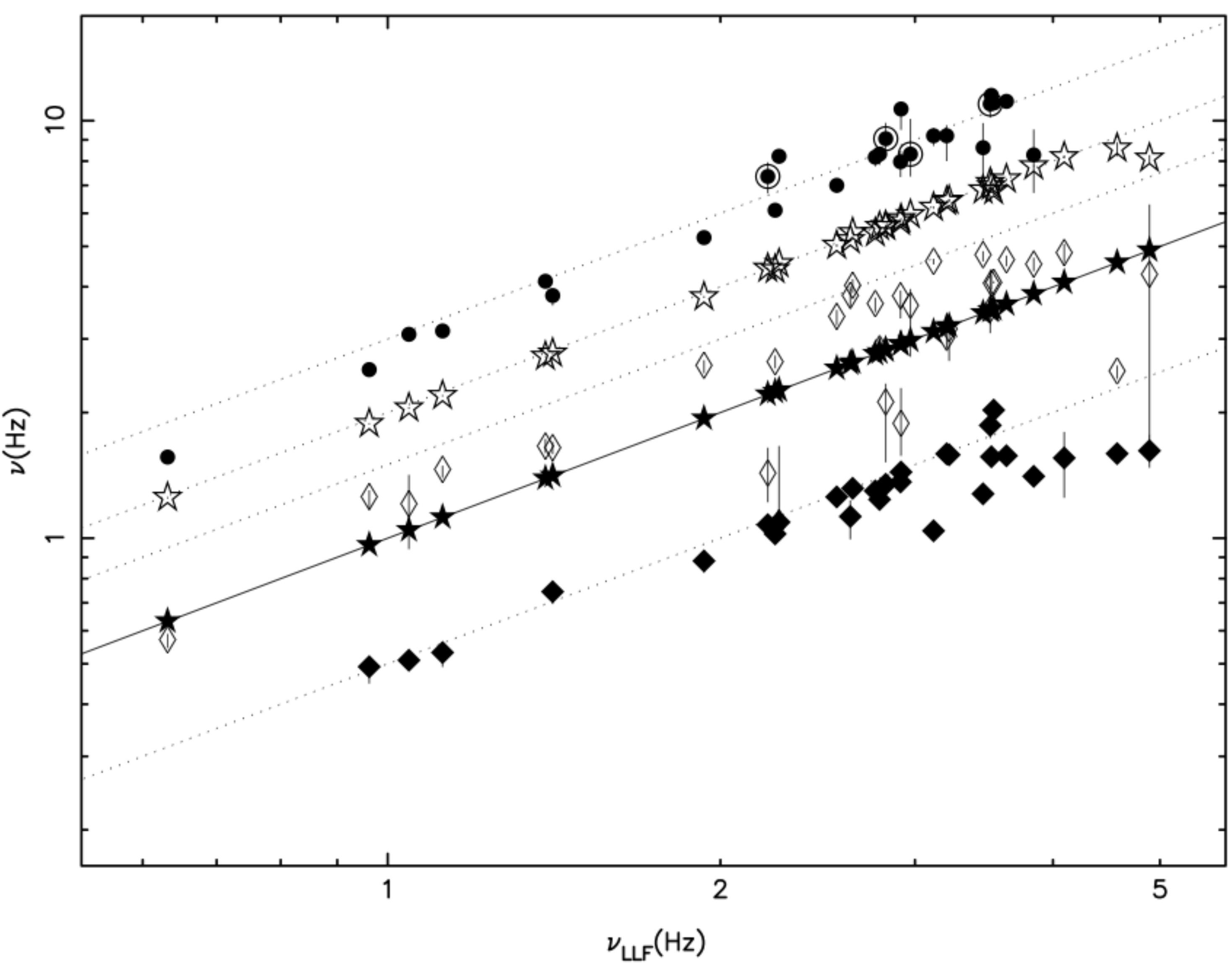} \\
\includegraphics[width=8.5cm] {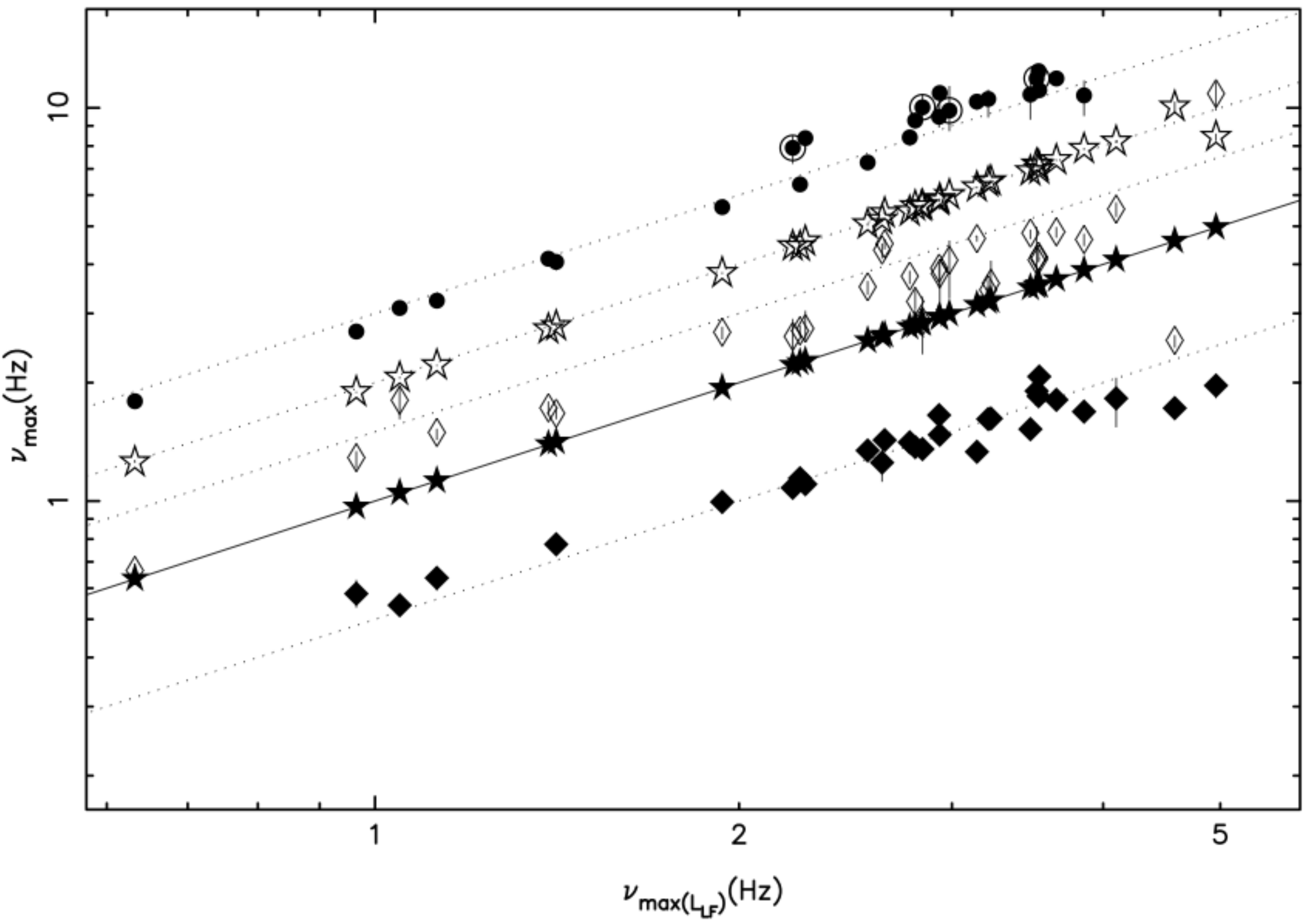} \\
\end{tabular}
\caption{ \label{Armoniche} Top panel: Central frequency of $L_{LF}$ (black stars), $L_{II}$ (empty stars), $L_{III}$ (black circles), $L_{1/2}$ (black diamonds) and $L_h$ (empty diamonds) versus the frequency of $L_{LF}$.  The black circles in an empty frame are 2 $\sigma$ detections of $L_{III}$. Lines of constant ratio R= 0.5, 1, 1.5, 2 and 3 are plotted: the solid line is R=1. Bottom panel: same as top panel, but the characteristic frequency $\nu_{max}=\sqrt{ \nu^2+(\Delta/2)^2}$ is plotted instead of $\nu$.}

\end{figure}

\subsection{Harmonic relations}
Figure \ref{Armoniche} (top panel) shows  the frequencies $\nu_{II}$, $\nu_{III}$, $\nu_{LF}$, $\nu_{I/2}$ and  $\nu_h$ plotted against the corresponding $\nu_{LF}$ (the data are reported in the tables from \ref{tabLLF} to \ref{tabLh}).  The harmonic relation between the $L_{LF}$ and $L_{II}$ is evident from the figure.  Quite some scattering is observed instead in the case of the broader components $L_{III}$ and $L_{1/2}$, with a  tendency of the points to lie slightly below the harmonic relation line.  Figure \ref{Armoniche} (bottom panel) suggests that the harmonic ratio is better represented for $L_{III}$ and $L_{1/2}$ if we consider, instead of the frequency of the components, their `characteristic frequency' $\nu_{max}=\sqrt{ \nu^2+(\Delta/2)^2}$ \citep{2002ApJ...572..392B}. This is a measure for the break frequency of a broad Lorentzian and around this frequency the component contributes most of its power per logarithmic frequency interval\footnote{Note that $\nu_{max}$ becomes equal to $\nu$ for narrow components.} \citep{2002ApJ...572..392B}. When considering $\nu$ (upper panel of Fig. \ref{Armoniche}), a fit to the $L_{III}$ and $L_{1/2}$ data-points with a line of constant ratio $\nu = A\,\nu_{LF}$ gives $A=2.91\pm0.02$ ($\chi^2$= 131.2, 24 d.o.f) and $A=0.455\pm0.004$ ($\chi^2$= 412.3, 27 d.o.f) for the two components respectively. The fitted slope is inconsistent with the expected harmonic ratios between the frequency of the each component and that of the $L_{LF}$ . If we consider $\nu_{max}$ instead (Fig. \ref{Armoniche} bottom panel), the fit gives $A=3.00\pm0.02$ for $L_{III}$ ($\chi^2$= 116.4, 24 d.o.f) and $A=0.49\pm0.004$ for $L_{1/2}$ ($\chi^2$= 191.4, 27 d.o.f).  In this case, the fitted slope is consistent with the expected harmonic ratios on the 1$\sigma$ level  for $L_{III}$ and on the 3$\sigma$ level for $L_{1/2}$. Although the fits are poor due to the scatter between the data-points, they support the visual impression given by Fig. \ref{Armoniche} that the harmonic ratio of the $L_{III}$ and $L_{1/2}$ is better represented when $\nu_{max}$ is used instead of $\nu$. 
 \newline
 The frequency of $L_h$ is included in Figure \ref{Armoniche} following a recent study of the PDS of XTE J1550-564 performed by \citet{2010ApJ...714.1065R}. The authors found that the $\nu_{max}$ of the component $L_h$ (named $L_{pn}$ by the authors) is in a 3/2 ratio with $\nu_{max(LF)}$, being at three times the characteristic frequency of the sub-harmonic $L_{1/2}$.  We find that, in the case of \source, $\nu_{max(h)}$ lies slightly below the line indicating a constant ratio R=3/2 with $\nu_{max(LF)}$ (Figure \ref{Armoniche} bottom panel). As $L_{h}$ is a rather broad component, this discrepancy is even stronger if we consider the centroid frequencies $\nu_{h}$ and $\nu_{LF}$ (Figure \ref{Armoniche} top panel).  

\section{Harmonics in a width-frequency plane}
\label{harm}

\begin{figure}
\centering
\includegraphics[width=8cm] {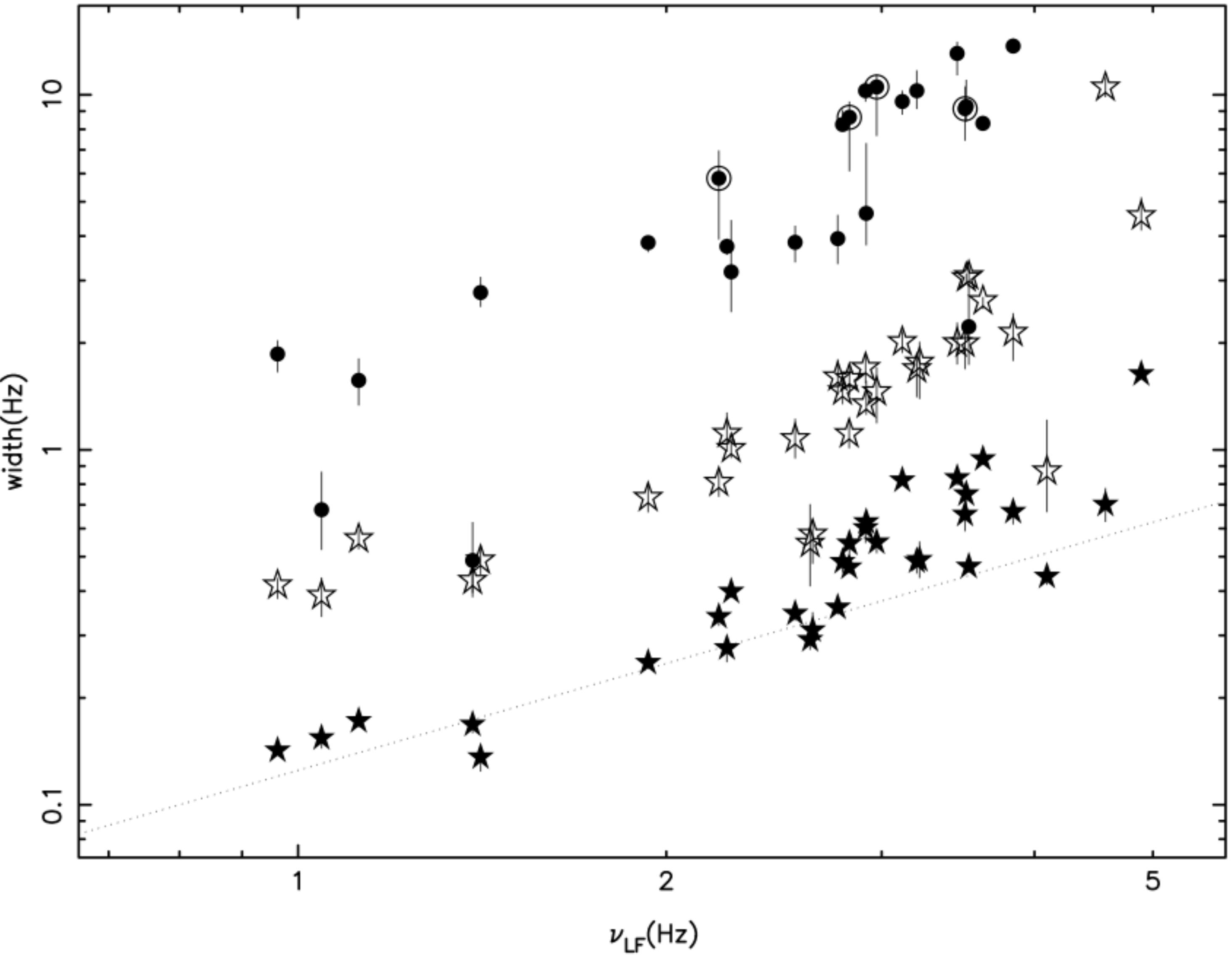}
\caption{ For each $L_{LF}$ (black stars), $L_{II}$ (empty stars) and $L_{III}$ (black dots) detection, the width is plotted against the centroid frequency of the $L_{LF}$ in the same PDS. The black dots in a white frame are 2 $\sigma$ detections consistent with $L_{III}$. The dashed line shows the slope of a constant quality factor ($Q$=10) trend.}  
\label{boh}
\end{figure}

\begin{figure*}
\centering
\includegraphics[width=16cm] {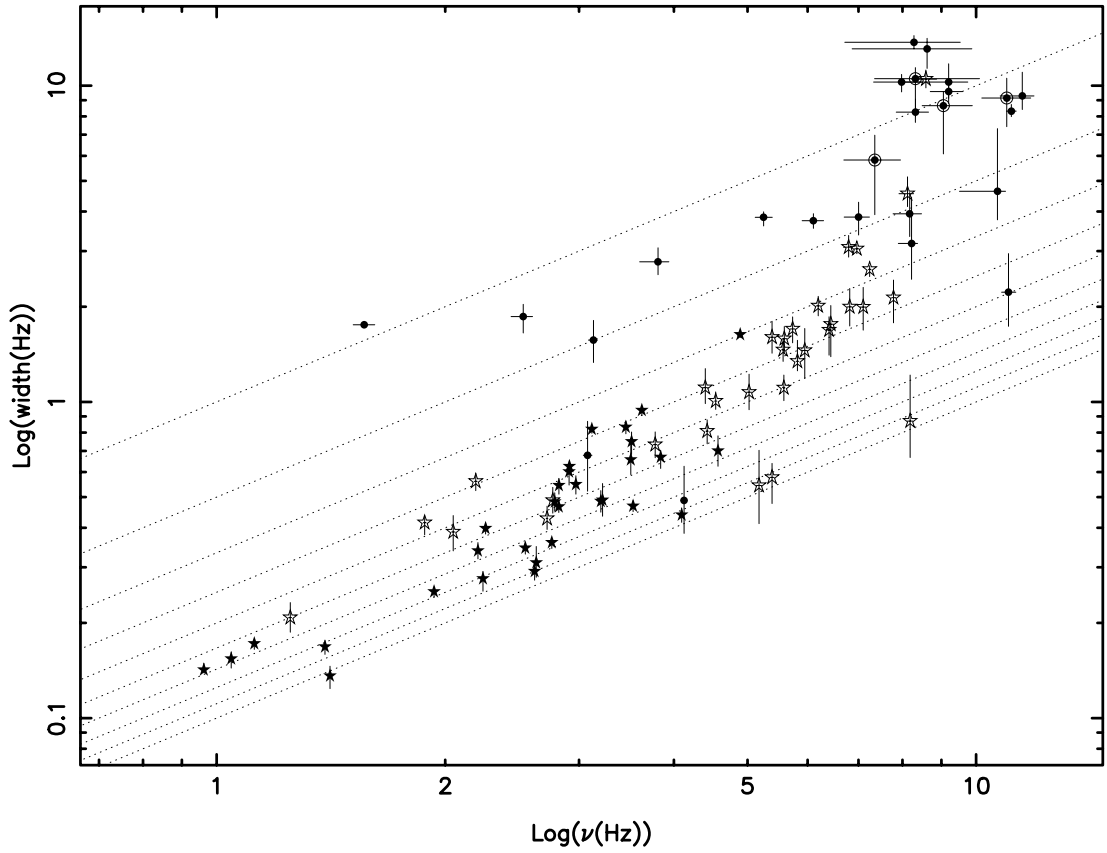}
\caption{ Harmonics on a width-frequency plane. Symbols are the same as in Figure \ref{boh}: for each $L_{LF}$ (black stars), $L_{II}$ (empty stars) and $L_{III}$ (black dots) component detected in the PDS, the width is plotted against the centroid frequency. The black dots in a white frame are 2 $\sigma$ detections consistent with $L_{III}$. Lines of constant quality factor $Q=\nu / \Delta$ are plotted (dashed lines, $Q$ increasing from 1 to 10 going from top to bottom). }  
\label{theplot}
\end{figure*}

\begin{figure*}
\centering
\hbox{ 
\hspace{0.01in}{(a)}
\includegraphics[width=8cm] {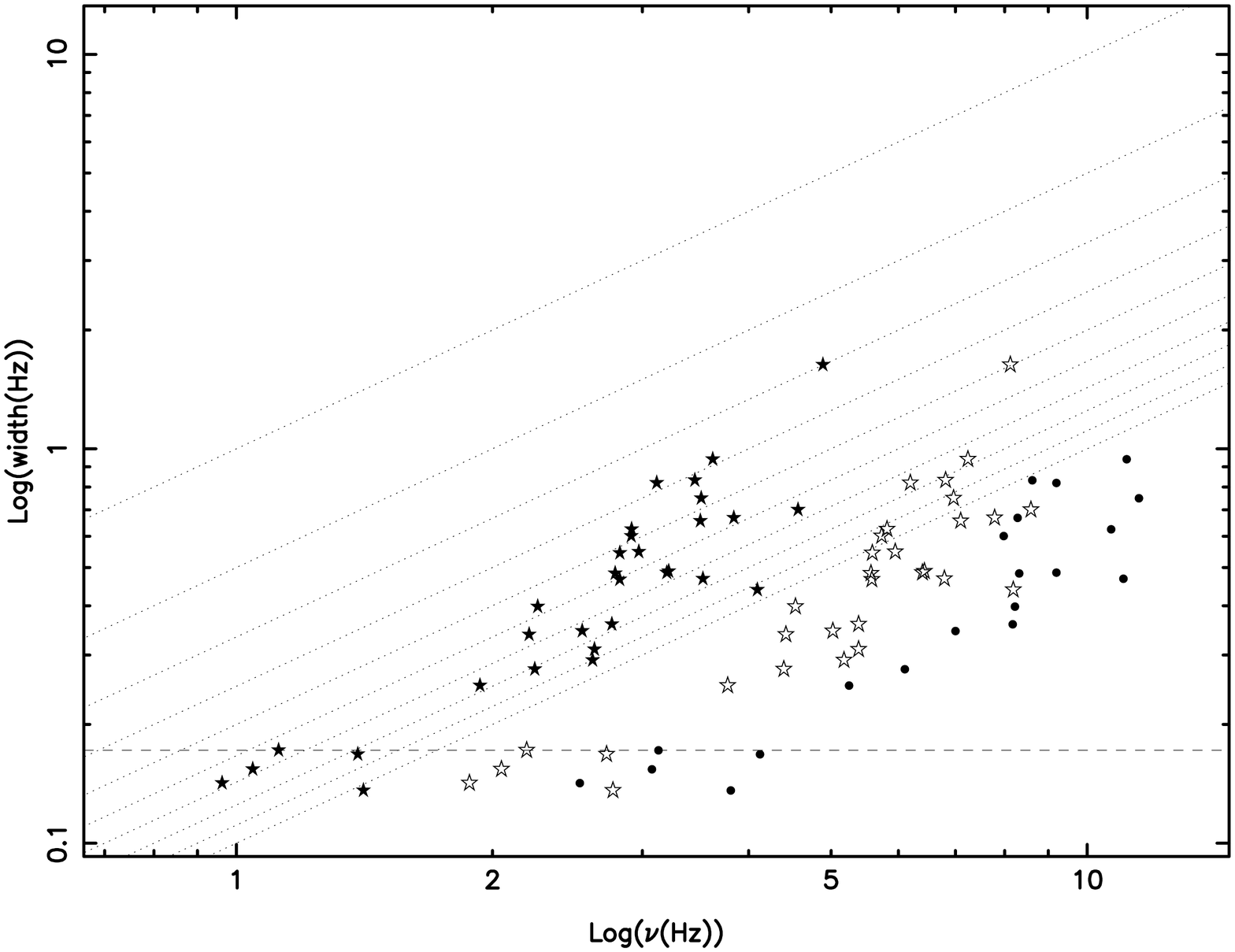}
\hspace{0.01in} {(b)}
\includegraphics[width=8cm] {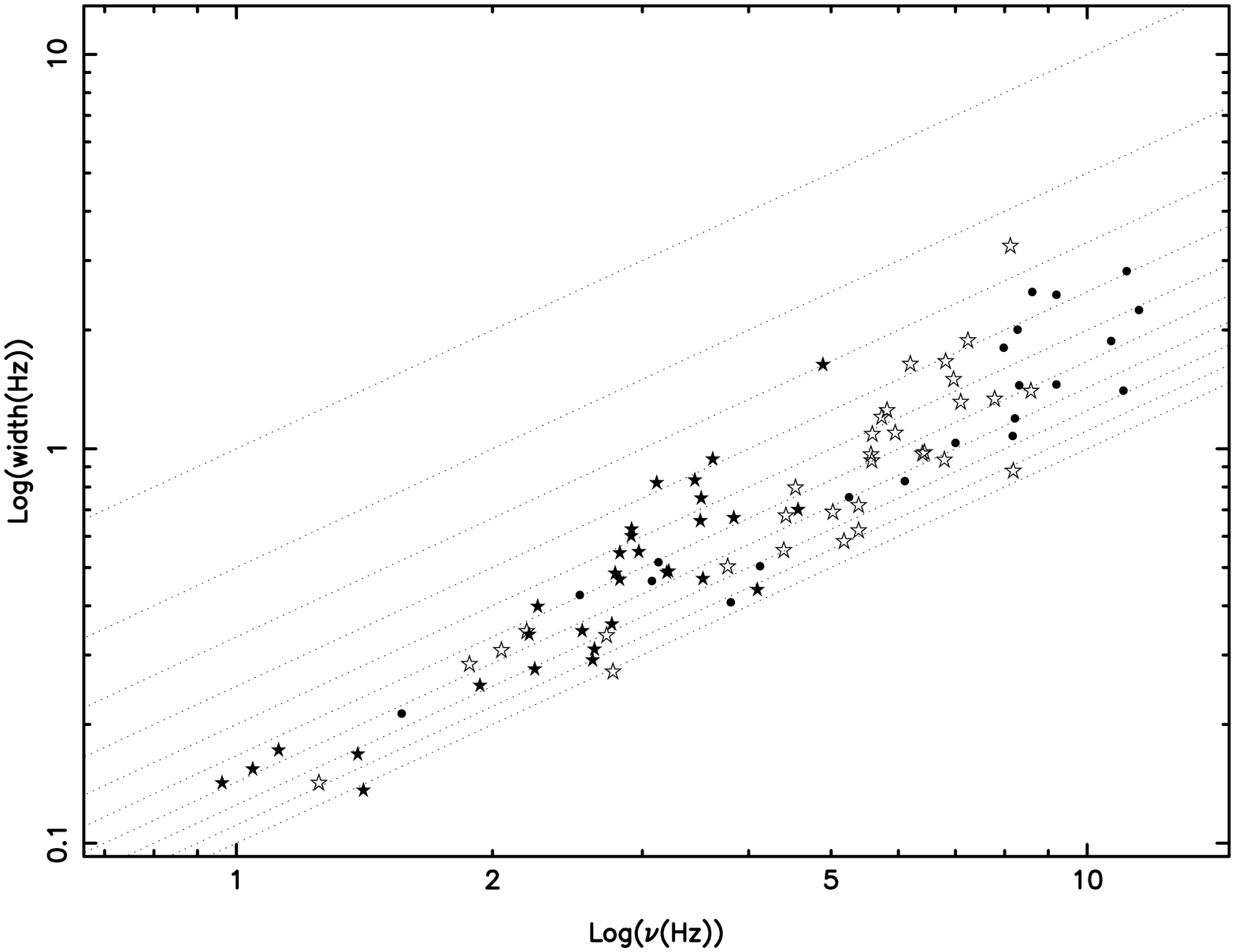}
}
\caption{Panels (a) and (b) show how the plot in Fig. \ref{theplot} would look like in case the harmonics origin from a signal modulated in amplitude and frequency respectively.  The points have the same frequency as the real data, but the width  of each harmonic is : (a)  the same as the width of the $L_{LF}$ in the same observation (amplitude modulation) (b)  calculated in order to obtain a constant $Q$ for all the harmonics from the same PDS (frequency modulation).}
\label{sym}
\end{figure*}

Fig. \ref{boh} shows the width of the $L_{LF}$ and its higher order harmonics $L_{II}$ and $L_{III}$ plotted against the centroid frequency of the $L_{LF}$. The three components describe three clear tracks on the diagram, and the width of each component grows with frequency. Moreover, it seems that the highest is the order of the harmonic, the broader the peak is. Nonetheless, a different interpretation arises when considering the plot of Fig.  \ref{theplot}, which provides a deeper insight on the behaviour of the harmonics. Here $L_{LF}$, $L_{II}$ and $L_{III}$  are considered again on a width-frequency plane, but the width $\Delta$ of each component is plotted against its own frequency $\nu$, without implying a relation with the $L_{LF}$ (the data are reported in the tables from \ref{tabLLF} to \ref{tabLh}). Lines of constant quality factor $Q$ are also plotted. Different symbols distinguish harmonics of different orders, with the same key used in Fig. \ref{boh}. 
With the exception of a few outliers, all points follow the same trend on this plot, where coherence decreases as centroid frequency increases. The harmonics can not be distinguished based on their width, as the tracks from different harmonics overlap one with another. Rather than depend on the order of the harmonic, a given $\Delta$ seems to be associated with a given frequency. A similar trend is observed if $\nu_{max}$ is plotted instead of $\nu$. \newline
The relation between the quality factors of the different harmonics of a non-sinusoidal signal is determined by the nature of the quasi-periodicity of the signal. The width-frequency plots in Fig. \ref{sym} have been produced by assuming the $\nu$ of each harmonic peak in our data and calculating the expected $\Delta$ in case of a purely frequency - or purely amplitude - modulated signal. 
For a purely amplitude modulation, the $\Delta$ of all the harmonic peaks would be the same as the amplitude of the $L_{LF}$.
On a frequency-width diagram that would result in a series of parallel tracks, one per each harmonic as shown  in Fig. \ref{sym} panel (a). The diagram is clearly different than that resulting from the real data in Fig. \ref{theplot}.  In the case of a frequency modulation, all the peaks have the same quality factor, that of the $L_{LF}$. This case is shown in Fig. \ref{sym} panel (b), which again is not consistent with the real data:  although a single track is formed, it remains parallel to the $Q=const$  lines, not reproducing the curved shape that appears in the real \source data set. A width-frequency diagram consistent with a pure frequency modulation was found instead for XTE J1550-564 \citep{2010ApJ...714.1065R}. 
\newline
The frequency-width diagram for components $L_{1/2}$ and $L_h$ is shown in Fig. \ref{nu-l-lh}. A comparison of Fig. \ref{nu-l-lh} and \ref{theplot} shows that  $L_{1/2}$ and $L_h$ do not share the behaviour of the other harmonics. 

\section{Discussion and Conclusions}
\label{disc}

We have analysed the PDS of \source during three {\it plateaux} states. The overall shape of the PDS is that typical of BHTs in their hard-intermediate state, with a strong low frequency QPO peak $L_{LF}$  superimposed to a band limited noise continuum (Fig. \ref{PDSesempi}, \ref{PDSmodel}). The spectra can be fitted with a  combination of Lorentzian components, several of which are in harmonic relation with $L_{LF}$. We detected up to two higher order harmonics ($L_{II}$ and $L_{III}$) forming an harmonic series of 1:2:3 with $L_{LF}$, plus the components $L_{1/2}$ and $L_h$ with frequency close to $\sim0.5$ and $\sim1.5$ times the frequency of the $L_{LF}$ respectively.  Few observations show residuals at frequency consistent to $\sim 4L_{LF}$, but a further Lorentzian component added at this frequency never results significant above 2$\sigma$.

\begin{figure}
\centering
\includegraphics[width=8.5cm] {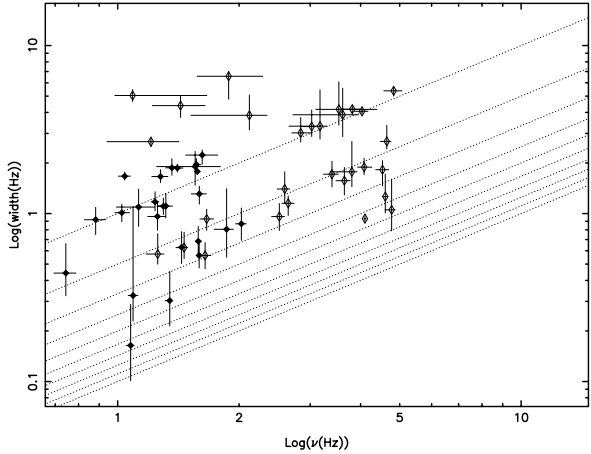}
\caption{ Width versus centroid frequency (same as Fig. \ref{theplot}) for the sub-harmonic $L_{1/2}$ (black diamonds) and the component $L_h$ (empty diamonds). Lines of constant quality factor $Q=\nu / \Delta$ are plotted as in Fig.  \ref{theplot} (dashed lines, $Q$ increasing from 1 to 10 from top to bottom) }  
\label{nu-l-lh}
\end{figure}

\noindent  In the case of  XTE J1550-564, \citet{2010ApJ...714.1065R} proposed $L_{1/2}$ as the fundamental frequency of an harmonic 1:2:3:4 series including $L_{1/2}$, $L_{LF}$, $L_{h}$ and  $L_{II}$. In a similar scenario for \source,  $L_{III}$ should also be included, leading to a 1:2:3:4:6 series. Nonetheless, there is no strong evidence of a 2/3 harmonic ratio between the frequency of the  $L_{LF}$ and of  $L_{h}$ in the case of \source, even when  the characteristic frequency of the components $\nu_{max}$ is considered in place of $\nu$. Moreover, the plot in Figure \ref{theplot} (which is not significantly affected by the choice of $\nu$ or $\nu_{max}$) indicates that $L_{LF}$ and its higher order harmonics are related beyond their frequency ratio as they broaden and lose coherence together as their frequency increases, while $L_{1/2}$ and $L_h$ behave differently (Fig. \ref{nu-l-lh}) as if they did not belong to the $L_{LF}$ harmonic series. We conclude that the $L_{1/2}$ is likely not the fundamental in the $L_{LF}$ harmonic series, i.e. is not included in the series itself. $L_{1/2}$ and  $L_{h}$ could instead be produced by a different phenomenon than that responsible for the $L_{LF}$. 

\noindent Although the plot in Fig.   \ref{theplot} evidences a relation between $L_{LF}$ and its higher order harmonics $L_{II}$ and $L_{III}$, it also rises doubts on the nature of this relation. The commonly accepted idea is that the $L_{LF}$ and the harmonics describe together the same signal from a quasi-periodic  oscillator, appearing as separated peaks in the PDS as a consequence  of the Fourier representation. However, it is not trivial to identify a signal modulation that is able to produce the trend in Fig.  \ref{theplot}. For example, we have shown that such a behavior cannot be reproduced by a simple combination of frequency or amplitude modulation. Moreover, Fig. \ref{Armoniche} shows that the frequency of $L_{III}$ across different observations is better distributed around an harmonic ratio of 3 with the frequency of $L_{LF}$ when considering $\nu_{max}$ instead of $\nu$ \footnote{Note that, because $L_{LH}$ is narrow, there is no significant difference between  $\nu_{max}$ and $\nu$ for this component. The same holds for  $L_{II}$. }. This is not what is usually expected in the context of harmonic decomposition, where the centroid $\nu$ is the frequency expected to be in harmonic ratio with the fundamental frequency. A broad component whose $\nu_{max}$ is in harmonic ratio with the fundamental frequency can not be regarded as a proper harmonic, but only as a signal whose characteristic frequency is consistent with an harmonic of the fundamental frequency.  The complexity of these results invites to consider a scenario involving more than one real oscillator, where a physical phenomenon is triggering oscillations at multiple frequencies in the accretion disk or in the corona, each one resulting in an harmonic component of the PDS. The trigger could be such that the life time (i.e. the coherence) of each oscillator is determined only by its own frequency, in agreement with the the trend in Fig. \ref{theplot}. This toy-model example gives a feeling of the new perspectives that would be opened in the interpretation of the PDS if the harmonics were proven to have a meaning beyond the Fourier representation of a single quasi-periodic signal. Either way, the interpretation of the width-frequency plane of \source promises to offer new insights on the QPO generation phenomenon in this system and in other XRBs.

\section*{Acknowledgments} \noindent TMB and SEM acknowledge support from grant PRIN INAF 2008. The research leading to these results has received funding from the European CommunityÕs Seventh Frame-work Programme (FP7/2007-2013) under grant agreement number ITN 215212 Black Hole Universe.

\begin{table}
\begin{center}
\begin{tabular}{lll}
\hline
Obs ID & $\nu_{LLF}$ & $\Delta_{LLF}$  \\
\hline
10408-01-22-01 & 2.762 $_{-0.006}^{+0.005}$ & 0.359$_{-0.015}^{+0.007}$ \\[3pt]
10408-01-22-02 & 2.549 $_{-0.006}^{+0.005}$ & 0.345$_{-0.015}^{+0.014}$ \\[3pt]
10408-01-23-00 & 3.535 $_{-0.007}^{+0.007}$ & 0.468$_{-0.018}^{+0.019}$ \\[3pt]
10408-01-24-00 & 2.242 $_{-0.010}^{+0.005}$ & 0.276$_{-0.024}^{+0.015}$ \\[3pt]
10408-01-25-00 & 1.121 $_{-0.003}^{+0.002}$ & 0.172$_{-0.007}^{+0.005}$ \\[3pt]
10408-01-27-00 & 0.632 $_{-0.002}^{+0.002}$ & 0.071$_{-0.004}^{+0.005}$ \\[3pt]
10408-01-28-00 & 0.962 $_{-0.002}^{+0.002}$ & 0.142$_{-0.003}^{+0.007}$ \\[3pt]
10408-01-29-00 & 1.933 $_{-0.004}^{+0.003}$ & 0.251$_{-0.010}^{+0.009}$ \\[3pt]
10408-01-30-00 & 4.891 $_{-0.010}^{+0.010}$ & 1.632$_{-0.034}^{+0.034}$ \\[3pt]
10408-01-31-00 & 4.095 $_{-0.007}^{+0.006}$ & 0.439$_{-0.023}^{+0.023}$ \\[3pt]
20402-01-04-00 & 4.572 $_{-0.025}^{+0.021}$ & 0.700$_{-0.074}^{+0.081}$ \\[3pt]
20402-01-05-00 & 2.822 $_{-0.005}^{+0.004}$ & 0.466$_{-0.019}^{+0.020}$ \\[3pt]
20402-01-07-00 & 3.119 $_{-0.006}^{+0.006}$ & 0.819$_{-0.023}^{+0.022}$ \\[3pt]
20402-01-08-00 & 3.843 $_{-0.010}^{+0.011}$ & 0.668$_{-0.052}^{+0.035}$ \\[3pt]
20402-01-08-01 & 3.458 $_{-0.011}^{+0.010}$ & 0.832$_{-0.039}^{+0.038}$ \\[3pt]
20402-01-09-00 & 2.823 $_{-0.007}^{+0.007}$ & 0.544$_{-0.033}^{+0.029}$ \\[3pt]
20402-01-10-00 & 2.914 $_{-0.006}^{+0.006}$ & 0.625$_{-0.024}^{+0.026}$ \\[3pt]
20402-01-11-00 & 2.911 $_{-0.009}^{+0.008}$ & 0.601$_{-0.055}^{+0.036}$ \\[3pt]
20402-01-12-00 & 2.787 $_{-0.007}^{+0.006}$ & 0.483$_{-0.032}^{+0.033}$ \\[3pt]
20402-01-13-00 & 3.630 $_{-0.008}^{+0.007}$ & 0.940$_{-0.035}^{+0.023}$ \\[3pt]
20402-01-14-00 & 3.518 $_{-0.016}^{+0.013}$ & 0.749$_{-0.067}^{+0.054}$ \\[3pt]
20402-01-15-00 & 2.260 $_{-0.004}^{+0.005}$ & 0.398$_{-0.016}^{+0.016}$ \\[3pt]
20402-01-16-00 & 2.972 $_{-0.009}^{+0.008}$ & 0.548$_{-0.037}^{+0.036}$ \\[3pt]
20402-01-18-00 & 3.222 $_{-0.011}^{+0.011}$ & 0.489$_{-0.054}^{+0.063}$ \\[3pt]
20402-01-19-00 & 2.208 $_{-0.005}^{+0.005}$ & 0.338$_{-0.019}^{+0.021}$ \\[3pt]
20402-01-20-00 & 3.205 $_{-0.007}^{+0.007}$ & 0.485$_{-0.037}^{+0.039}$ \\[3pt]
20402-01-21-00 & 3.510 $_{-0.013}^{+0.012}$ & 0.656$_{-0.067}^{+0.109}$ \\[3pt]
20402-01-49-00 & 2.622 $_{-0.007}^{+0.006}$ & 0.291$_{-0.018}^{+0.019}$ \\[3pt]
20402-01-49-01 & 2.635 $_{-0.014}^{+0.011}$ & 0.310$_{-0.031}^{+0.039}$ \\[3pt]
20402-01-50-01 & 1.045 $_{-0.004}^{+0.002}$ & 0.154$_{-0.010}^{+0.008}$ \\[3pt]
20402-01-51-00 & 1.389 $_{-0.003}^{+0.002}$ & 0.168$_{-0.009}^{+0.008}$ \\[3pt]
20402-01-52-00 & 1.410 $_{-0.005}^{+0.003}$ & 0.136$_{-0.012}^{+0.010}$ \\[3pt]
\end{tabular}
\caption{ \label{tabLLF} Frequency  $\nu_{LLF}$, width $\Delta_{LLF}$ of the low frequency QPO $L_{LF}$ from the best fit to the PDS. }
\end{center}
\end{table}

\begin{table}
\begin{center}
\begin{tabular}{lll}
\hline
Obs ID & $\nu_{II}$ & $\Delta_{II}$  \\
\hline
10408-01-22-01 & 5.386 $_{-0.057}^{+0.051}$ & 1.601$_{-0.175}^{+0.198}$ \\[3pt]
10408-01-22-02 & 5.024 $_{-0.038}^{+0.036}$ & 1.073$_{-0.128}^{+0.151}$ \\[3pt]
10408-01-23-00 & 6.794 $_{-0.106}^{+0.079}$ & 3.087$_{-0.217}^{+0.273}$ \\[3pt]
10408-01-24-00 & 4.399 $_{-0.035}^{+0.025}$ & 1.110$_{-0.122}^{+0.164}$ \\[3pt]
10408-01-25-00 & 2.194 $_{-0.010}^{+0.009}$ & 0.561$_{-0.037}^{+0.035}$ \\[3pt]
10408-01-27-00 & 1.250 $_{-0.006}^{+0.005}$ & 0.208$_{-0.021}^{+0.024}$ \\[3pt]
10408-01-28-00 & 1.879 $_{-0.008}^{+0.008}$ & 0.415$_{-0.035}^{+0.020}$ \\[3pt]
10408-01-29-00 & 3.780 $_{-0.021}^{+0.020}$ & 0.733$_{-0.067}^{+0.071}$ \\[3pt]
10408-01-30-00 & 8.121 $_{-0.206}^{+0.154}$ & 4.555$_{-0.408}^{+0.596}$ \\[3pt]
10408-01-31-00 & 8.186 $_{-0.067}^{+0.066}$ & 0.870$_{-0.203}^{+0.346}$ \\[3pt]
20402-01-04-00 & 8.586 $_{-0.358}^{+0.286}$ & 10.492$_{-0.650}^{+0.717}$ \\[3pt]
20402-01-05-00 & 5.582 $_{-0.010}^{+0.019}$ & 1.108$_{-0.099}^{+0.108}$ \\[3pt]
20402-01-07-00 & 6.195 $_{-0.037}^{+0.039}$ & 2.012$_{-0.140}^{+0.145}$ \\[3pt]
20402-01-08-00 & 7.784 $_{-0.043}^{+0.051}$ & 2.137$_{-0.358}^{+0.291}$ \\[3pt]
20402-01-08-01 & 6.817 $_{-0.047}^{+0.048}$ & 1.998$_{-0.260}^{+0.286}$ \\[3pt]
20402-01-09-00 & 5.590 $_{-0.031}^{+0.039}$ & 1.585$_{-0.160}^{+0.146}$ \\[3pt]
20402-01-10-00 & 5.818 $_{-0.023}^{+0.022}$ & 1.344$_{-0.090}^{+0.222}$ \\[3pt]
20402-01-11-00 & 5.733 $_{-0.038}^{+0.041}$ & 1.702$_{-0.164}^{+0.155}$ \\[3pt]
20402-01-12-00 & 5.570 $_{-0.026}^{+0.026}$ & 1.462$_{-0.119}^{+0.115}$ \\[3pt]
20402-01-13-00 & 7.242 $_{-0.030}^{+0.031}$ & 2.626$_{-0.134}^{+0.069}$ \\[3pt]
20402-01-14-00 & 6.966 $_{-0.049}^{+0.046}$ & 3.049$_{-0.107}^{+0.095}$ \\[3pt]
20402-01-15-00 & 4.541 $_{-0.016}^{+0.015}$ & 1.006$_{-0.059}^{+0.070}$ \\[3pt]
20402-01-16-00 & 5.949 $_{-0.057}^{+0.046}$ & 1.454$_{-0.267}^{+0.253}$ \\[3pt]
20402-01-18-00 & 6.437 $_{-0.053}^{+0.056}$ & 1.761$_{-0.371}^{+0.257}$ \\[3pt]
20402-01-19-00 & 4.424 $_{-0.015}^{+0.015}$ & 0.808$_{-0.071}^{+0.069}$ \\[3pt]
20402-01-20-00 & 6.403 $_{-0.035}^{+0.034}$ & 1.688$_{-0.283}^{+0.171}$ \\[3pt]
20402-01-21-00 & 7.100 $_{-0.065}^{+0.061}$ & 1.995$_{-0.309}^{+0.307}$ \\[3pt]
20402-01-49-00 & 5.178 $_{-0.043}^{+0.040}$ & 0.544$_{-0.132}^{+0.159}$ \\[3pt]
20402-01-49-01 & 5.387 $_{-0.038}^{+0.035}$ & 0.577$_{-0.100}^{+0.062}$ \\[3pt]
20402-01-50-01 & 2.049 $_{-0.010}^{+0.011}$ & 0.388$_{-0.050}^{+0.049}$ \\[3pt]
20402-01-51-00 & 2.723 $_{-0.008}^{+0.007}$ & 0.428$_{-0.033}^{+0.040}$ \\[3pt]
20402-01-52-00 & 2.771 $_{-0.013}^{+0.012}$ & 0.488$_{-0.043}^{+0.049}$ \\[3pt]
\end{tabular}
\caption{ \label{tabLII} Frequency  $\nu_{II}$, width $\Delta_{II}$ of the harmonic QPO $L_{II}$ from the best fit to the PDS. }
\end{center}
\end{table}

\begin{table}
\begin{center}
\begin{tabular}{lll}
\hline
Obs ID & $\nu_{III}$ & $\Delta_{III}$  \\
\hline
10408-01-22-01 & 8.176 $_{-0.401}^{+0.298}$ & 3.936$_{-0.603}^{+0.654}$ \\[3pt]
10408-01-22-02 & 7.002 $_{-0.299}^{+0.234}$ & 3.841$_{-0.470}^{+0.439}$ \\[3pt]
10408-01-23-00 & 11.030 $_{-0.228}^{+0.249}$ & 2.224$_{-0.490}^{+0.721}$ \\[3pt]
10408-01-24-00 & 6.105 $_{-0.204}^{+0.192}$ & 3.742$_{-0.204}^{+0.197}$ \\[3pt]
10408-01-25-00 & 3.135 $_{-0.048}^{+0.037}$ & 1.569$_{-0.236}^{+0.241}$ \\[3pt]
10408-01-27-00 & 1.564 $_{-0.051}^{+0.051}$ & 1.754$_{-0.021}^{+0.039}$ \\[3pt]
10408-01-28-00 & 2.533 $_{-0.091}^{+0.073}$ & 1.861$_{-0.209}^{+0.174}$ \\[3pt]
10408-01-29-00 & 5.250 $_{-0.132}^{+0.140}$ & 3.835$_{-0.233}^{+0.157}$ \\[3pt]
20402-01-07-00 & 9.201 $_{-0.500}^{+0.399}$ & 9.584$_{-0.790}^{+0.698}$ \\[3pt]
20402-01-08-00 & 8.282 $_{-1.567}^{+1.247}$ & 13.721$_{-0.642}^{+0.685}$ \\[3pt]
20402-01-08-01 & 8.618 $_{-1.752}^{+1.253}$ & 13.078$_{-1.734}^{+1.037}$ \\[3pt]
20402-01-10-00 & 10.667 $_{-1.159}^{+0.265}$ & 4.635$_{-0.869}^{+2.678}$ \\[3pt]
20402-01-11-00 & 7.977 $_{-0.649}^{+0.501}$ & 10.268$_{-0.711}^{+0.590}$ \\[3pt]
20402-01-12-00 & 8.321 $_{-0.473}^{+0.334}$ & 8.251$_{-0.269}^{+0.826}$ \\[3pt]
20402-01-13-00 & 11.128 $_{-0.165}^{+0.161}$ & 8.312$_{-0.325}^{+0.411}$ \\[3pt]
20402-01-14-00 & 11.501 $_{-0.745}^{+0.417}$ & 9.279$_{-0.869}^{+1.751}$ \\[3pt]
20402-01-15-00 & 8.221 $_{-0.322}^{+0.152}$ & 3.170$_{-0.728}^{+1.269}$ \\[3pt]
20402-01-20-00 & 9.202 $_{-1.198}^{+0.548}$ & 10.267$_{-1.148}^{+1.458}$ \\[3pt]
20402-01-50-01 & 3.079 $_{-0.040}^{+0.041}$ & 0.678$_{-0.156}^{+0.190}$ \\[3pt]
20402-01-51-00 & 4.126 $_{-0.040}^{+0.037}$ & 0.488$_{-0.104}^{+0.138}$ \\[3pt]
20402-01-52-00 & 3.812 $_{-0.206}^{+0.127}$ & 2.773$_{-0.247}^{+0.299}$ \\[3pt]
20402-01-09-00$^\dag$ & 9.054 $_{-0.564}^{+0.829}$ & 8.648$_{-2.56}^{+0.935}$ \\[3pt]
20402-01-16-00$^\dag$ & 8.318 $_{-0.969}^{+1.784}$ & 10.529$_{-2.877}^{+0.299}$ \\[3pt]
20402-01-19-00$^\dag$ & 7.355 $_{-0.659}^{+0.596}$ & 5.821$_{-1.912}^{+0.299}$ \\[3pt]
20402-01-21-00$^\dag$ & 10.971 $_{-0.799}^{+0.830}$ & 9.143$_{-1.725}^{+1.414}$ \\[3pt]
\end{tabular}
\caption{ \label{tabLIII} Frequency  $\nu_{III}$, width $\Delta_{III}$ of the harmonic QPO $L_{III}$ from the best fit to the PDS. \newline $^\dag$ 2 sigma detection.}
\end{center}
\end{table}


\begin{table}
\begin{center}
\begin{tabular}{lll}
\hline
Obs ID & $\nu_{1/2}$ & $\Delta_{1/2}$  \\
\hline
10408-01-22-01 & 1.298 $_{-0.042}^{+0.042}$ & 1.105$_{-0.119}^{+0.129}$ \\[3pt]
10408-01-22-02 & 1.255 $_{-0.068}^{+0.057}$ & 0.961$_{-0.164}^{+0.207}$ \\[3pt]
10408-01-23-00 & 2.026 $_{-0.066}^{+0.057}$ & 0.869$_{-0.177}^{+0.209}$ \\[3pt]
10408-01-24-00 & 1.024 $_{-0.040}^{+0.046}$ & 1.012$_{-0.117}^{+0.123}$ \\[3pt]
10408-01-25-00 & 0.532 $_{-0.041}^{+0.028}$ & 0.703$_{-0.106}^{+0.135}$ \\[3pt]
10408-01-28-00 & 0.492 $_{-0.044}^{+0.024}$ & 0.618$_{-0.098}^{+0.168}$ \\[3pt]
10408-01-29-00 & 0.882 $_{-0.055}^{+0.051}$ & 0.919$_{-0.169}^{+0.170}$ \\[3pt]
10408-01-30-00 & 1.620 $_{-0.146}^{+0.151}$ & 2.229$_{-0.266}^{+0.159}$ \\[3pt]
10408-01-31-00 & 1.556 $_{-0.307}^{+0.241}$ & 1.904$_{-0.430}^{+0.440}$ \\[3pt]
20402-01-04-00 & 1.594 $_{-0.072}^{+0.065}$ & 1.309$_{-0.171}^{+0.146}$ \\[3pt]
20402-01-07-00 & 1.041 $_{-0.037}^{+0.034}$ & 1.669$_{-0.086}^{+0.098}$ \\[3pt]
20402-01-08-00 & 1.406 $_{-0.046}^{+0.046}$ & 1.866$_{-0.098}^{+0.104}$ \\[3pt]
20402-01-08-01 & 1.277 $_{-0.062}^{+0.055}$ & 1.661$_{-0.134}^{+0.163}$ \\[3pt]
20402-01-09-00 & 1.346 $_{-0.035}^{+0.024}$ & 0.303$_{-0.089}^{+0.150}$ \\[3pt]
20402-01-10-00 & 1.440 $_{-0.029}^{+0.030}$ & 0.630$_{-0.126}^{+0.150}$ \\[3pt]
20402-01-11-00 & 1.363 $_{-0.047}^{+0.039}$ & 1.872$_{-0.184}^{+0.247}$ \\[3pt]
20402-01-12-00 & 1.238 $_{-0.027}^{+0.026}$ & 1.175$_{-0.138}^{+0.174}$ \\[3pt]
20402-01-13-00 & 1.575 $_{-0.021}^{+0.021}$ & 1.780$_{-0.063}^{+0.061}$ \\[3pt]
20402-01-14-00 & 1.567 $_{-0.024}^{+0.040}$ & 1.956$_{-0.143}^{+0.167}$ \\[3pt]
20402-01-15-00 & 1.092 $_{-0.030}^{+0.030}$ & 0.325$_{-0.096}^{+0.760}$ \\[3pt]
20402-01-18-00 & 1.584 $_{-0.025}^{+0.027}$ & 0.684$_{-0.119}^{+0.154}$ \\[3pt]
20402-01-19-00 & 1.077 $_{-0.027}^{+0.022}$ & 0.164$_{-0.063}^{+0.124}$ \\[3pt]
20402-01-20-00 & 1.592 $_{-0.022}^{+0.022}$ & 0.563$_{-0.089}^{+0.112}$ \\[3pt]
20402-01-21-00 & 1.862 $_{-0.128}^{+0.073}$ & 0.806$_{-0.257}^{+0.601}$ \\[3pt]
20402-01-49-00 & 1.127 $_{-0.134}^{+0.105}$ & 1.093$_{-0.254}^{+0.304}$ \\[3pt]
20402-01-49-01 & 1.316 $_{-0.064}^{+0.052}$ & 1.108$_{-0.167}^{+0.209}$ \\[3pt]
20402-01-50-01 & 0.510 $_{-0.024}^{+0.021}$ & 0.370$_{-0.118}^{+0.118}$ \\[3pt]
20402-01-52-00 & 0.744 $_{-0.049}^{+0.044}$ & 0.442$_{-0.118}^{+0.221}$ \\[3pt]
\end{tabular}
\caption{ \label{tabL1/2} Frequency  $\nu_{1/2}$ and width $\Delta_{1/2}$ of the Lorentzian component $L_{1/2}$, from the best fit to the PDS. }
\end{center}
\end{table}

\clearpage
\begin{table}
\begin{center}
\begin{tabular}{lll}
\hline
Obs ID & $\nu_{h}$ & $\Delta_{h}$  \\
\hline
10408-01-22-01 & 3.647 $_{-0.169}^{+0.122}$ & 1.569$_{-0.278}^{+0.313}$ \\[3pt]
10408-01-22-02 & 3.398 $_{-0.162}^{+0.121}$ & 1.715$_{-0.275}^{+0.333}$ \\[3pt]
10408-01-23-00 & 4.101 $_{-0.025}^{+0.025}$ & 0.932$_{-0.037}^{+0.035}$ \\[3pt]
10408-01-24-00 & 2.647 $_{-0.095}^{+0.092}$ & 1.151$_{-0.176}^{+0.120}$ \\[3pt]
10408-01-25-00 & 1.461 $_{-0.066}^{+0.030}$ & 0.627$_{-0.088}^{+0.147}$ \\[3pt]
10408-01-27-00 & 0.571 $_{-0.025}^{+0.019}$ & 0.686$_{-0.077}^{+0.089}$ \\[3pt]
10408-01-28-00 & 1.257 $_{-0.081}^{+0.045}$ & 0.574$_{-0.074}^{+0.184}$ \\[3pt]
10408-01-29-00 & 2.595 $_{-0.111}^{+0.075}$ & 1.400$_{-0.240}^{+0.370}$ \\[3pt]
10408-01-30-00 & 4.288 $_{-2.607}^{+2.013}$ & 19.979$_{-0.052}^{+0.013}$ \\[3pt]
10408-01-31-00 & 4.831 $_{-0.263}^{+0.237}$ & 5.377$_{-0.369}^{+0.242}$ \\[3pt]
20402-01-04-00 & 2.516 $_{-0.109}^{+0.076}$ & 0.959$_{-0.167}^{+0.188}$ \\[3pt]
20402-01-07-00 & 4.606 $_{-0.080}^{+0.065}$ & 1.261$_{-0.253}^{+0.455}$ \\[3pt]
20402-01-08-00 & 4.533 $_{-0.173}^{+0.163}$ & 1.824$_{-0.364}^{+0.241}$ \\[3pt]
20402-01-08-01 & 4.776 $_{-0.173}^{+0.095}$ & 1.052$_{-0.263}^{+0.549}$ \\[3pt]
20402-01-09-00 & 2.121 $_{-0.602}^{+0.224}$ & 3.846$_{-0.692}^{+1.230}$ \\[3pt]
20402-01-10-00 & 1.884 $_{-0.310}^{+0.404}$ & 6.567$_{-1.776}^{+0.442}$ \\[3pt]
20402-01-11-00 & 3.806 $_{-0.447}^{+0.114}$ & 1.772$_{-0.328}^{+0.917}$ \\[3pt]
20402-01-12-00 & 2.842 $_{-0.140}^{+0.172}$ & 3.020$_{-0.357}^{+0.709}$ \\[3pt]
20402-01-13-00 & 4.646 $_{-0.168}^{+0.109}$ & 2.688$_{-0.265}^{+0.662}$ \\[3pt]
20402-01-14-00 & 4.082 $_{-0.150}^{+0.186}$ & 1.889$_{-0.188}^{+0.240}$ \\[3pt]
20402-01-15-00 & 1.089 $_{-0.105}^{+0.574}$ & 5.042$_{-0.399}^{+0.435}$ \\[3pt]
20402-01-16-00 & 3.612 $_{-0.894}^{+0.340}$ & 3.871$_{-1.003}^{+1.698}$ \\[3pt]
20402-01-18-00 & 3.173 $_{-0.515}^{+0.130}$ & 3.314$_{-0.540}^{+2.131}$ \\[3pt]
20402-01-19-00 & 1.432 $_{-0.213}^{+0.215}$ & 4.392$_{-0.659}^{+0.617}$ \\[3pt]
20402-01-20-00 & 3.026 $_{-0.183}^{+0.082}$ & 3.295$_{-0.423}^{+0.840}$ \\[3pt]
20402-01-21-00 & 3.533 $_{-0.435}^{+0.860}$ & 4.169$_{-0.810}^{+1.912}$ \\[3pt]
20402-01-49-00 & 3.815 $_{-0.231}^{+0.190}$ & 4.179$_{-0.197}^{+0.182}$ \\[3pt]
20402-01-49-01 & 4.033 $_{-0.168}^{+0.138}$ & 4.061$_{-0.136}^{+0.120}$ \\[3pt]
20402-01-50-01 & 1.209 $_{-0.268}^{+0.209}$ & 2.679$_{-0.180}^{+0.118}$ \\[3pt]
20402-01-51-00 & 1.662 $_{-0.063}^{+0.067}$ & 0.928$_{-0.134}^{+0.132}$ \\[3pt]
20402-01-52-00 & 1.647 $_{-0.058}^{+0.056}$ & 0.563$_{-0.094}^{+0.109}$ \\[3pt]
\end{tabular}
\caption{ \label{tabLh} Frequency  $\nu_{h}$ and width $\Delta_{h}$ of the Lorentzian component $L_{h}$, from the best fit to the PDS. }
\end{center}
\end{table}

\bibliographystyle{mn}
\bibliography{1915rev2_Last}

\begin{thebibliography}{25}
\expandafter\ifx\csname natexlab\endcsname\relax\def\natexlab#1{#1}\fi

\bibitem[{Belloni(2010)}]{Bel10}
Belloni T., 2010, in Lecture Notes in Physics, Vol. 794, The Jet Paradigm,
  Belloni T., ed., Springer Berlin / Heidelberg, pp. 53--84

\bibitem[{{Belloni} \& {Hasinger}(1990)}]{1990A&A...227L..33B}
{Belloni} T., {Hasinger} G., 1990, \aap, 227, L33

\bibitem[{{Belloni} {et~al.}(2000){Belloni}, {Klein-Wolt}, {M{\'e}ndez}, {van
  der Klis}, \& {van Paradijs}}]{2000A&A...355..271B}
{Belloni} T., {Klein-Wolt} M., {M{\'e}ndez} M., {van der Klis} M., {van
  Paradijs} J., 2000, \aap, 355, 271

\bibitem[{{Belloni} {et~al.}(1997){Belloni}, {Mendez}, {King}, {van der Klis},
  \& {van Paradijs}}]{1997ApJ...488L.109B}
{Belloni} T., {Mendez} M., {King} A.~R., {van der Klis} M., {van Paradijs} J.,
  1997, \apjl, 488, L109+

\bibitem[{{Belloni} {et~al.}(2002){Belloni}, {Psaltis}, \& {van der
  Klis}}]{2002ApJ...572..392B}
{Belloni} T., {Psaltis} D., {van der Klis} M., 2002, \apj, 572, 392

\bibitem[{{Castro-Tirado} {et~al.}(1992){Castro-Tirado}, {Brandt}, \&
  {Lund}}]{1992IAUC.5590....2C}
{Castro-Tirado} A.~J., {Brandt} S., {Lund} N., 1992, \iaucirc, 5590, 2

\bibitem[{{Fender} \& {Belloni}(2004)}]{2004ARA&A..42..317F}
{Fender} R., {Belloni} T., 2004, \araa, 42, 317

\bibitem[{{Fender} {et~al.}(1999){Fender}, {Garrington}, {McKay}, {Muxlow},
  {Pooley}, {Spencer}, {Stirling}, \& {Waltman}}]{1999MNRAS.304..865F}
{Fender} R.~P., {Garrington} S.~T., {McKay} D.~J., {Muxlow} T.~W.~B., {Pooley}
  G.~G., {Spencer} R.~E., {Stirling} A.~M., {Waltman} E.~B., 1999, \mnras, 304,
  865

\bibitem[{{Greiner} {et~al.}(2001{\natexlab{a}}){Greiner}, {Cuby}, \&
  {McCaughrean}}]{2001Natur.414..522G}
{Greiner} J., {Cuby} J.~G., {McCaughrean} M.~J., 2001{\natexlab{a}}, \nat, 414,
  522

\bibitem[{{Greiner} {et~al.}(2001{\natexlab{b}}){Greiner}, {Cuby},
  {McCaughrean}, {Castro-Tirado}, \& {Mennickent}}]{2001A&A...373L..37G}
{Greiner} J., {Cuby} J.~G., {McCaughrean} M.~J., {Castro-Tirado} A.~J.,
  {Mennickent} R.~E., 2001{\natexlab{b}}, \aap, 373, L37

\bibitem[{{Greiner} {et~al.}(1996){Greiner}, {Morgan}, \&
  {Remillard}}]{1996ApJ...473L.107G}
{Greiner} J., {Morgan} E.~H., {Remillard} R.~A., 1996, \apjl, 473, L107+

\bibitem[{{Ingram} \& {Done}(2011)}]{2011MNRAS.415.2323I}
{Ingram} A., {Done} C., 2011, \mnras, 415, 2323

\bibitem[{{Leahy} {et~al.}(1983){Leahy}, {Darbro}, {Elsner}, {Weisskopf},
  {Kahn}, {Sutherland}, \& {Grindlay}}]{1983ApJ...266..160L}
{Leahy} D.~A., {Darbro} W., {Elsner} R.~F., {Weisskopf} M.~C., {Kahn} S.,
  {Sutherland} P.~G., {Grindlay} J.~E., 1983, \apj, 266, 160

\bibitem[{{Markwardt} {et~al.}(1999){Markwardt}, {Swank}, \&
  {Taam}}]{1999ApJ...513L..37M}
{Markwardt} C.~B., {Swank} J.~H., {Taam} R.~E., 1999, \apjl, 513, L37

\bibitem[{{Psaltis}(2006)}]{2006csxs.book....1P}
{Psaltis} D., 2006, in Compact stellar X-ray sources, {Lewin} W.~H.~G., {van
  der Klis} M., eds., pp. 1--38

\bibitem[{{Rao} {et~al.}(2010){Rao}, {Belloni}, {Stella}, {Zhang}, \&
  {Li}}]{2010ApJ...714.1065R}
{Rao} F., {Belloni} T., {Stella} L., {Zhang} S.~N., {Li} T., 2010, \apj, 714,
  1065

\bibitem[{{Reig} {et~al.}(2000){Reig}, {Belloni}, {van der Klis}, {M{\'e}ndez},
  {Kylafis}, \& {Ford}}]{2000ApJ...541..883R}
{Reig} P., {Belloni} T., {van der Klis} M., {M{\'e}ndez} M., {Kylafis} N.~D.,
  {Ford} E.~C., 2000, \apj, 541, 883

\bibitem[{{Remillard} \& {McClintock}(2006)}]{Rem06}
{Remillard} R.~A., {McClintock} J.~E., 2006, \araa, 44, 49

\bibitem[{{Rodriguez} \& {Varni{\`e}re}(2011)}]{2011arXiv1104.3865R}
{Rodriguez} J., {Varni{\`e}re} P., 2011, ArXiv e-prints

\bibitem[{{Sobczak} {et~al.}(2000){Sobczak}, {McClintock}, {Remillard}, {Cui},
  {Levine}, {Morgan}, {Orosz}, \& {Bailyn}}]{2000ApJ...544..993S}
{Sobczak} G.~J., {McClintock} J.~E., {Remillard} R.~A., {Cui} W., {Levine}
  A.~M., {Morgan} E.~H., {Orosz} J.~A., {Bailyn} C.~D., 2000, \apj, 544, 993

\bibitem[{{Trudolyubov}(2001)}]{2001ApJ...558..276T}
{Trudolyubov} S.~P., 2001, \apj, 558, 276

\bibitem[{{van der Klis}(1989)}]{1989ARA&A..27..517V}
{van der Klis} M., 1989, \araa, 27, 517

\bibitem[{{van der Klis}(2005)}]{2005AN....326..798V}
---, 2005, Astronomische Nachrichten, 326, 798

\bibitem[{{van der Klis}(2006)}]{2006csxs.book..39V}
{van der Klis} R., 2006, Compact stellar X-ray sources, 39

\bibitem[{{Zhang} {et~al.}(1995){Zhang}, {Jahoda}, {Swank}, {Morgan}, \&
  {Giles}}]{1995ApJ...449..930Z}
{Zhang} W., {Jahoda} K., {Swank} J.~H., {Morgan} E.~H., {Giles} A.~B., 1995,
  \apj, 449, 930

\end{thebibliography}

\end{document}